\documentclass[preprint,12pt]{aastex}
\def\etal {{\it et~al.}}
\def\ang   {\AA\/}

\def\msun{M_{\odot}}

\def\mdot{\dot M}
\begin{document}

\doublespace

\title{Simultaneous ASCA and HST/GHRS
observations of
Cygnus~X-2/V1341~Cygni}

\author{S. D. Vrtilek, J. C. Raymond}
\affil{Harvard-Smithsonian Center for Astrophysics, 60 Garden St., 
Cambridge MA 02138. svrtilek@cfa.harvard.edu,
jraymond@cfa.harvard.edu}

\author{B. Boroson}
\affil{MIT, Cambridge, MA 02138.  bboroson@space.mit.edu}

\author{R. McCray}
\affil{JILA, University of Colorado, Boulder, CO 80309.
DICK@jila.colorado.edu}

\author{A. Smale, T. Kallman} 
\affil{NASA/Goddard Space Flight Center, Greenbelt, MD 20177.
alan@osiris.gsfc.nasa.gov, tim@xstar.gsfc.nasa.gov}

\author{F. Nagase}
\affil{ISAS, 3-1-1, Yoshimodai, Sagamihara, Kanagawa 229-8510. 
nagase@astro.isas.ac.jp}
.
\newpage
.
\section{Abstract}
We present results from ultraviolet and
X-ray observations of the low mass X-ray binary Cygnus X-2.  
The simultaneous HST/GHRS and ASCA observations took place 
during the low state
of an 82-day cycle.  
We compare
our observations as well as archival IUE and RXTE data 
with models that
predict ultraviolet and optical continuum emission from  
an X-ray heated disk 
and
a Roche-lobe-filling star.
The model predictions are consistent with observed optical,
ultraviolet, and X-ray variations over both orbital and long-term
periods.  
The X-ray spectral state, the luminosities
implied by fits to the X-ray data, the ultraviolet continuum
and line fluxes, and the mass accretion
rates obtained from fits to the ultraviolet continuum are 
consistent with
location of our observations
on the normal and horizontal branches of the Z-shaped
X-ray color-color diagram.
A combination of changes to mass accretion rate and obstruction
by a warped disk can be invoked as a possible 
explanation for the motion of the ``Z'' in the color-color plane.

The GHRS/G160M measurements concentrated on 
NV~($\lambda1238.8;~\lambda1242.8$)
and HeII~($\lambda 1640.5$). 
The low-resolution (GHRS/G140L) observations captured
SiIV ($\lambda~1393.8;~\lambda1402.8$), NIV~($\lambda~1486.5$),
and CIV~($\lambda~1548$); 
absorption lines detected in the spectra are
interstellar.
Although the relative line fluxes
are consistent with emission from an X-ray heated accretion
disk corona,
predictions from
models of line emission from simple disks do not fit
the observed emission line profiles.  The lack of double peaks
suggests that most of the line emission is from the surface of the
companion and the radial velocities 
(80-130 km s$^{-1}$) 
are consistent with emission from the optical star at the orbital phase
(0.70-0.74) of our observations.

stars: neutron -- X-rays: binaries 
--X-rays: individual (Cygnus X-2)

\newpage

\section{Introduction}

Cygnus X-2 \footnote{Based partially on observations with the NASA/ESA
 Hubble Space Telescope, obtained at the
 Space Telescope Science Institute, which is
 operated by the Association of Universities
 for Research in Astronomy, Inc. under NASA
 contract No. NAS5-26555.}
is a low mass X-ray binary (LMXB) with an orbital period 
of 9.843 days (Cowley, Crampton, \& Hutchings 1979).  
Its optical companion, V1341 Cygni, is a 0.7$\msun$ star of 
15th magnitude with spectral type varying from A5 to F2.  
RXTE observations have shown the presence of a long-term period
between 78-82 days
in Cyg X-2 which has since been confirmed with data from earlier missions
(Wijnands, Kuulkers, \& Smale 1996; Paul, Kitamoto, \& Makino 2000).

Cyg X-2 is a ``Z-source": an LMXB which 
shows three distinct X-ray spectral
states, referred to as the flaring, normal, and horizontal branches
(FB, NB, and HB), which form a `Z' shape
on an X-ray color-color diagram.
While Cyg X-2 displays no high-Q pulsations, each state shows
distinct modes of quasi-periodic oscillations
(QPOs) in X-ray emission.
Cyg X-2 is the only ``Z-class source''
that regularly displays all three X-ray spectral states (FB, NB, and HB)
with their associated QPOs
and is bright in the ultraviolet (Hasinger \etal~1990; 
Vrtilek \etal~1990 [hereafter Paper I]).
It is also among the Z-sources that show significant motion of the Z pattern
on the color-color plane (Wijnands~\etal~1997).

Simultaneous ultraviolet and X-ray observations of LMXBs 
show a unique relationship
between the X-ray spectral states and the ultraviolet flux.
Sco X-1
and Cyg X-2, the two brightest Z-sources, have ultraviolet flux that 
is directly 
correlated with
the X-ray spectral state of the source (Paper I; Vrtilek \etal~1991a);
there is a factor-of-three increase in ultraviolet flux from the HB to the FB.
Most of the optical and ultraviolet
luminosity ($L_{opt} \sim L_{ultraviolet} \sim 10^{-2} L_X$) likely 
comes from the
accretion flow as it is illuminated by the X-rays from the neutron
star.
The ultraviolet spectra of Cyg X-2 and Sco X-1 (Paper I; Vrtilek~\etal~1991a)
are dominated by strong emission lines of
N\thinspace V~$\lambda~1240$, C\thinspace IV~$\lambda~1550$, and
He\thinspace II~$\lambda~1640$.
The ultraviolet line strengths, ratios, and profiles vary noticeably ($\sim
20\%$) on the shortest timescales ($\sim 30$min for Sco X-1
and $\sim 120$min for Cyg X-2) that can be probed
with IUE. 
There is evidence that the X-ray lines in Cyg X-2 vary in the same way
as the ultraviolet lines in that both are
enhanced during the
FB (Vrtilek \etal~1991a,b).

Here we present simultaneous ultraviolet and X-ray observations of
Cyg X-2 taken with the GHRS on HST 
and the GIS and SIS detectors on ASCA.
With HST we can for the first time
measure variability of the
ultraviolet spectra of LMXBs on short ($50\,$ms $\le \Delta t \le
30\,$min) timescales.
This enables a search for an ultraviolet manifestation of the QPOs
observed in X-rays. 
Finding QPOs in the ultraviolet would be important: whereas the
X-ray flux can be easily modulated by geometric effects
and emission in the visible can be confused with that of the
non-collapsed secondary, the ultraviolet observations are most likely to
yield direct evidence of phenomena related to the
accretion disk.
In addition the HST spectral resolution, $\Delta V \sim 19$km~s$^{-1}$ 
(HRS G160M), enables
the dissection of accretion flows according to velocity, a
measurement of local
physical conditions and dynamics that the X-ray observations
cannot do.
From delays
between X-ray and ultraviolet line variations, 
the radial structure of velocity in the disk can be analyzed.
This provides an
unprecedented probe into the structure and dynamics of the
accretion flow, enabling us to test models of the ultraviolet emission and
constrain properties of the secondary star.

In this paper we compare the simultaneous HST/GHRS and ASCA data as 
well as archival IUE 
data with models
predicting ultraviolet continuum emission from the X-ray heated disk and 
star.  We interpret the double-peaked long-term period of Cyg X-2 
in terms of a saddle shaped disk. 
The ultraviolet spectral features are tested 
with simple models for line emission from a disk and from an X-ray
heated corona above the disk.   A comparison is made of the ultraviolet 
spectra of Cyg X-2 
with those of similar binary systems taken with the GHRS. 
The observations and analysis are
presented in section 3 and our interpretation is discussed in 
section 4.

\section{Observations and Analysis}

Figure 1 shows the location of our observation in comparison with the 
one-day averaged lightcurves obtained from the All Sky Monitor (ASM) on 
RXTE
provided by the RXTE/ASM team.  
A description of the RXTE can be found in Levine~\etal~(1996).
Since the ASM data currently available cover a considerably longer 
time than those used
by Wijnands, Kuulkers, \& Smale (1996) we 
re-determined the long-term period using 
an Analysis of Variance (ANOVA) method with 
(F-statistic) as described in Davies (1990;1991). 
For the data shown in Figure 1 our most significant period is 82.5$\pm$2.5 days;
using the entire RXTE data set to date (through April 2003) we find 81.7$\pm$0.6
days;  
we use the 81.7 day period in this paper.
The 82-day cycle has two maxima and two minima; we define phase zero as
the center of the deepest minimum.
Figure 1c shows the 82-day lightcurve we determine 
and Figure 1b the lightcurve
superposed on the data. 
This indicates that our 
simultaneous HST and ASCA observations occurred during one 
of the low states 
of the 82-day cycle. 

The ASCA lightcurves and HST coverage are plotted in Figure 2.  
The HST/GHRS took eight separate observations on 1995 Dec. 7 from 
3:26 UT (JD~2,450,058.643) to
1995 Dec. 7 12:49~UT (JD 2,450,059.034) which covers 
the orbital phases 0.70-0.74.
ASCA observed Cyg X-2 from 1995 Dec. 6 22:15~UT (JD 2,450,058.427) 
to 1995 Dec. 7 22:01~UT (JD 2,450,059.417) covering the orbital
phase range 0.68-0.78.  
The orbital phases were calculated using an ephemeris of JD2443161.68 
for phase 0.0 and a period of 9.843 d (Cowley, Crampton, \& Hutchings 
1979), where phase 0.0 refers to 
superior conjunction of the optical star.

The first four HST observations were taken with the GHRS G160M (17~km~s$^{-1}$) grating
centered alternately on 1240\AA~(N~V) and 1640\AA~(He~II) in the 
RAPID mode with
a time resolution of 0.5s (Figure 3).   
A description of the GHRS and 
gratings is given
by Cardelli, Savage, \& Ebbets (1990). 
The GHRS observations were
taken after the installation of COSTAR and the repairs to the
faulty electronics of side 1.
The last four observations used the GHRS/G140L (140 km s$^{-1}$)
in the wavelength region 1220-1550\AA~in 
order to test for ultraviolet continuum
and line strength variability (Figure 4). 
The G140L was used in RAPID mode with a time resolution of 50ms to
search for the quasi-periodic oscillations so far seen only
in the X-rays.
 Our use of the GHRS in the RAPID mode
prevented us from over-sampling the spectrum, checking for bad counts,
and correcting for the Doppler shift due to the spacecraft orbit.
In addition, the continua for the G160M exposures are poorly determined
because most of the counts ($\sim 80$\%) are due to background.
Hence the standard products produced for counts per diode per time
interval are not suited for temporal variability studies.  

The X-ray countrates from both the GIS and SIS detectors on ASCA
(described in Tanaka, Inoue, \& Holt [1994])
varied by less than 10\% during our
observation.
Consequently the X-ray color-color diagram covers only a small
segment of the `Z' (Figure 5).  The slope, when compared to the
complete Z shown
by Hasinger \etal~(1990), suggests that we are on the HB,
possibly at the turning point into the NB.
No significant dips in X-ray countrate were visible during our
observation.

Much of our SIS data were taken in FAST mode which presents problems for
spectral analysis.  FAST mode complicates the removal of 
flickering pixels.  Also, the response depends on position and there is no
position information in FAST mode.  We obtained the FAST mode data for
the timing analysis (search for QPOs).
Our SIS BRIGHT mode data of Cyg X-2 suffer from pile-up because
the count rate is so high.
Pileup arises when more than one photon is collected in the same pixel (or
neighboring pixels) in the SIS CCDs within an integration time (4 seconds
for BRIGHT mode.)  The result of pileup is that multiple input photons are
not detected separately, but a signal is recorded corresponding 
to the sum of the
energies of the photons.  Thus pileup can falsely increase the high-energy
part of the spectrum while decreasing the low-energy spectrum.  (The
latter effect is usually negligible, as the counts peak at low energies.)

Our treatment of pileup follows that of Ebisawa \etal~(1996).
For analysis of the SIS BRIGHT mode spectra, we
discard the central circle of 13 pixel radius, as this is the region in
which pileup is most severe.  Then we construct a model of the high energy
spectrum produced by piled-up photons using Ebisawa's algorithm; this model
is subtracted from the total counts before fitting.  
Piled-up counts contribute
10\% at 5~keV
and 30\% at 8~keV, according to Ebisawa's analysis.  With the
central 13 pixel radius region removed, the count rate is 86~cts~s$^{-1}$
with a piled up rate of 1.5~cts~s$^{-1}$ for A1-3 (A4 had no BRIGHT
mode data); without the central
region removed the count rate is 
140~cts~s$^{-1}$ and the
piled up count rate is 4.6~cts~s$^{-1}$. 
Once the model for the pileup is taken into account, the SIS BRIGHT mode
spectra agree with the GIS models (detailed below) for E~$>$~1 keV.
Below 1~keV pilup effects are too strong to correct.

\subsection{Continuum Fits}

We used basic parameters for the source as found in Cowley, Crampton, 
and Hutchings (1979), McClintock~\etal~(1984) and paper I: a distance 
to the source of 8 Kpc; an E$_{B-V}$ of 0.45;
and a 0.7 $\msun$ F star companion with an
effective temperature of 7,000K.
We assume that the 82-day period is due to a warping
and precession of the disk induced by the effects of uneven X-ray
irradiation on the disk.
This
is a commonly accepted means for producing superorbital periods in
systems such as Her X-1, LMC X-4, and SMC X-1
(Clarkson~\etal~2002; 2003; Cheng, Vrtilek,
\& Raymond 1995; Vrtilek \etal~1997).

\subsubsection{Ultraviolet: HST and IUE}

\subsubsubsection{Disk Model} 

We simulate the warp with 
a saddle-shape for the inner disk of Cyg X-2 (Figure 6),
in order to reproduce the double-peaked shape of the long-term 
X-ray lightcurves. 
In effect the shape of the disk is an ``inversion'' of the 
average X-ray lightcurve.
To determine the inversion, we loop through 28,800 points (120 points in latitude
and 240 points in longitude) on the neutron star surface.
This corresponds to a grid coarseness of 0.025 r$_{NS}$;
refining the grid by a factor of two in each dimension
gives the same result within errors, so this number of
grid points should be sufficient.
For each point, the program computes analytically whether the
line towards the viewer intersects the disk.
We follow Howarth \& Wilson (1983a,b) in that the coordinates are
fixed to the disk so that it is the line-of-sight that changes with
the long-term phase.  This coordinate system allows us to specify
the shape of the disk in a frame in which it does not vary over the
long-term phase.
The equation of the
saddle-shaped
disk is given by
$z_d = \pi\theta_w(y_d^2 - x_d^2)/180r_d$ for (r$_d < 1.7 r_{NS}$);
it is constrained to
revert to a simple disk ($z_d$~=~0) in the orbital plane for large
radii (r$_d \ge 1.7 r_{NS}$).

The parameters describing the disk, listed in Table~1, are the 
radius of maximum 
disk warp, R$_{mw}$; the disk thickness, $\theta_d$, in degrees; 
the tilt of the disk from the orbital plane, 
$\alpha_d$;  
and the line-of-nodes parameter, $\Delta \psi$ (this parameter is
``invisible" in the sense that it does not
count against the $\chi^2$ but sets the
phase origin of the model relative to the
observed long-term X-ray light curve).  
Our best-fit parameters are a maximum saddle
height of 1.3r$_{NS}$ at a distance of 1.7r$_{NS}$ radius with
a tilt of the inner disk from the orbital plane of
$\alpha_d$=4.1$\pm0.4$ degrees at long-term phase 0.5 (Table 1).
The tilt of the inner disk from the orbital plane allows for the
uneven depths of the lightcurve:  at long-term phase 0.0 the disk
is tilted away from the observer causing a deeper dip than at
long-term phase 0.5.
The X-ray flux we see, as determined by the amount of X-rays
that are hidden by the saddle-shaped inner disk,  
is compared to the RXTE ASM data
in Figure 7. This is also the flux seen by the outer disk and
the companion star.

\subsubsubsection{X-ray reprocessing}

We adapt our model for determining continuum flux from X-ray reprocessing
on the companion star and disk as described in Paper I in several ways: 
(1) We replace the spherical shape assumed for the companion
star in Paper I with a tidally distorted
shape that represents the filled Roche lobe of V1341 Cygni
as was done for Her X-1 by Cheng, Vrtilek,
\& Raymond (1995).
(2) 
Instead of computing the temperature due to X-ray heating on each
of 40 points on the companion star surface we computed the temperature
at 7200 points (120 in longitude and 60 in latitude). 
(3) The disk is represented in 3-dimensions by using 1000 annular rings 
divided into viewing angles determined by a 15 by 29 array for each
of 8 ``quadrants''. 

Given the shape of the disk and a distorted Roche-lobe filling
companion star we calculate the flux
over the range 1150-7427$\AA$.
At each point on the companion star and the disk 
a temperature due to X-ray
heating is computed; the contribution to the flux is determined
by summing spectra from the library of observed stellar spectra
(Cannizzo \& Kenyon 1987) for a star at the computed temperature. 

Predictions of the model from the ultraviolet
and optical U and V bands are shown in Figure 8.  
The U and V optical predictions are compared to published photometry
of Cowley, Crampton, and Hutchings (1979) and Chevalier, Bonazzola,
and Ilovaisky~(1976).
The extremes of U and V magnitudes reported are given as dotted lines 
in Figure 8.
We have used archival
IUE data for comparison with the ultraviolet predictions.
Since the GHRS had a limited continuum coverage (Fig. 4) we integrated 
over
the wavelength regions 1340-1380$\AA$ and 1420-1480$\AA$ to
compare the UV fluxes.
For the optical V band we integrated over the range 5000-6000 $\AA$
and for the U band over 3510-3520$\AA$ (although the U band is 
centered on 3500$\AA$ the stellar library [Cannizzo \& Kenyon 1987]
that we use to determine
fluxes has
no values between 3200-3500$\AA$). 
We then converted the computed fluxes to magnitudes using
formulae taken from Lang (1980).

Because the 82-day period has a large uncertainty the long term phase of the
optical and IUE observations cannot be determined.
The predicted U magnitudes fully encompass the range observed; 
the predicted V magnitudes overlap with, but are on average greater
than, the reported extremes.
The ultraviolet data---including
the current observations and all archival IUE observations---fall within 
extremes
given by a lower limit to $\mdot$ of
3.2~$\times~10^{-9}\msun$ yr$^{-1}$ and an upper limit of
1.8~$\times~10^{-8}\msun$ yr$^{-1}$, consistent with the results of
Paper I.
There is uncertainty in the flux level, both for the model and the data.
The HST data are
calibrated to only 10\% in absolute flux, and there is a difference in
resolution between IUE and GHRS.  As for the model, uncertainties in 
reddening and stellar temperature can shift the flux (see the Appendix
in Paper I for a discussion of these uncertainties).  The relative
errors in flux are likely to be much smaller than the 
uncertainty in overall flux
normalization.
The best fit to the HST continuum data (shown as a filled triangle)
corresponds
to an $\mdot$ of 6.0~$\times~$10$^{-9}\msun$ yr$^{-1}$;
this is consistent with a 
location in the HB (Paper I)  
(Figure 4; the thin solid lines represent the models). 
The $\mdot$ values computed by applying this model to 
the G160M data (observations H2 and H3; {Figure 3) were somewhat higher.  
Either we caught 
the source at the
upper bend of the `Z' where the HB connects with the NB or the
very small continuum band available in the G160M is not sufficient
to constrain the model properly and the $\mdot$ values are not
reliable (here the background is much greater than the continuum).

\subsubsection{X-ray: ASCA}

The X-ray spectra observed with the ASCA GIS were modelled with 
the sum of  
Comptonized bremsstrahlung (following Sunyaev \& Titarchuk 1980), 
a blackbody, and an emission line at $\approx$
1 keV (Table 2).  In Figure 9 we show the spectra from the ASCA
observations divided into three intervals:  A1-A4; A6-18; and A9-A11; 
an observation of Cyg X-2 taken during ASCA's performance verification phase is 
also depicted (Smale \etal~1994).  The region around 2.2 keV is
excluded due to an instrumental effect not fully accounted for in the
calibration (Mukai, personal communication 1996).  The spectral fits
are consistent with interval A6-A8 located on the upper NB, the PV
phase observation on the HB, and intervals A1-A4 and A9-A11 in a 
transition from the HB to the NB.  Specifically the blackbody component
is reduced in the HB and line emission is enhanced in the NB; we note
that a blackbody component is present in Cyg X-2 at all phases of the
Z with the {\it exception} of the upper HB (Kuulkers, van der Klis, 
\& van Paradijs~1995).  The
luminosities implied by the above spectral fits for an assumed distance
of 8 kpc ranges from (3-5) $\times 10^{37}$ ergs~s$^{-1}$ consistent with
the HB and the upper NB as inferred by the ultraviolet continuum fits. 
The reality of the emission features around 1 keV have been questioned
by Psaltis, Lamb, \& Miller ~(1995; PLM95) who are able to fit ASCA data of 
Cygnus X-2
with a Comptonized thermal emission model that does not require spectral
features. 
They suggest that features near
1 keV may be due to known problems in
Comptonization models.  An approximation to the PLM95 model assumes
that below a given energy (taken as a free parameter) the spectrum is a
blackbody, while above that energy it has a Comptonized shape, with
temperature equal to that of the blackbody.  We find that fits of 
these models are
not as good as the Comptonized + blackbody + line models; 
e.g., 
$\chi_{\nu}^2$=1.42 instead of 0.995 for A1-A4 and 
$\chi_{\nu}^2$=1.73 instead of 1.22
for A9-11.  
The transition energy between the blackbody
and Comptonized components occurs at a low energy (~0.6 keV).
Line features around 1 keV have been reported for Cyg X-2 by several observers:
Vrtilek \etal~1986, 1988 reported lines in Einstein OGS and SSS spectra
of Cyg X-2, attributing a line at 0.96~keV with Fe~XX and Ni~XX and
a line at 1.12~keV with Fe~XXIII-XXIV.
The presence of these features was confirmed by Kuulkers~\etal~(1997)
using the Low Energy Concentrator Spectrometer on-board {\it Beppo}SAX.
In addition, 
Branduardi-Raymont~\etal~(1984) using {\it Ariel}~V Experiment C,
Chiappetti~\etal~(1990) using the EXOSAT CMA, Lum~\etal~(1992)
using the {\it Einstein} Focal Plane Crystal Spectrometer, and
Smale~\etal~(1993; 1994) using the Broad Band X-ray Telescope and the 
ASCA SIS all report evidence for excess emission near 1~keV
from Cyg X-2 (cf. Kuulkers~\etal~1997).

\subsection{Ultraviolet Spectral Features}

Table 3 gives the line fluxes and velocities for the G160M and G140L
observations shown in Figures 3 and 4.  The velocities are determined 
from flux-weighted wavelengths and are consistent with those observed
in optical (He II 4686) by Cowley, 
Crampton, \& Hutchings (1979) for the orbital
phase of our observations (0.70-0.74).
The error in the fluxes include both counting statistics and  
uncertainties introduced by 
choosing background and wavelength boundaries. 
No significant change is observed during the different
observations, which is not surprising since they cover a very small 
fraction of the
binary phase space.  The 
N~V doublet is clearly resolved with the doublet ratio close to 
1.2:1 rather
than the 2:1 expected from thin accretion disks.  A 
similar result was found from
GHRS observations of Sco X-1 (A N~V doublet ratio of 1.1:1; 
Kallman, Boroson, \& Vrtilek 1998) and  
the narrow lines seen in Her X-1 (ratio~1.2:1; Boroson \etal~1996).
The Cyg X-2 N~V lines can also be fit by including a broad emission
component (Figure 10) such as was observed in Her X-1; 
but this is not required.  
For the C~IV line we were only able to get an upper limit to the line 
flux, as the line is cut off by the edge of the detector.  
For this
line, the line velocity is found by flux-weighting the wavelengths 
only near the peak, not over the entire line profile (where the
flux is greater than the background) as for the other lines.  

In Figure 11 we show a composite of the four GHRS G140L spectra of 
Cyg X-2 together with those of
Sco X-1 (Kallman, Boroson, \& Vrtilek 1998) and Her X-1 (Boroson
\etal~1997).  In each case the solid line depicts the best-fit X-ray heated
accretion disk model for the continuum.  
(The slope in the Her X-1 continuum spectrum is due to the strong
contribution from the star; for Cyg X-2 and Sco X-1 the continuum is
dominated by X-ray heating of the disk).
While the saddle-shaped disk provides good fits to the continuum
emission, the intensities of  
many of the spectral features remain unaccounted for by
photoionization models, such as the XSTAR code (Kallman \&\ McCray 1982;
Kallman \& Krolik 1993).
For both Sco X-1 and Her X-1 the O\,V line at 1371 \AA\ is surprisingly
strong compared to
N\,V, contrary to theoretical
expectations (Raymond 1993, Ko \&\ Kallman 1994),
whereas Cyg X-2 showed no O~V $\lambda$1371
and relatively
strong N~IV $\lambda$1486. Since Sco X-1 and Cyg X-2 are both
LMXBs of
the same type the
differences in their ultraviolet spectra and the similarity of the Sco X-1 and
Her X-1 ultraviolet spectra
is surprising.
The N\,V line in Her~X-1 (not shown)  
is much stronger than the C\,IV line suggesting abundance anomalies
(Raymond 1993).
Cyg X-2 and Her X-1 have similar mass accretion rates  
when Cyg X-2 is in the HB.
The difference in flux can be attributed to reddening and distance
effects.
The presence of N~IV], its strength relative to other lines, 
and the absence of O~V in Cyg X-2 are consistent with the predictions 
of Raymond (1993) model F.  Since N~IV] is an intercombination line
which requires somewhat lower densities it is produced 
preferentially at the outer edges of the disk where densities
are lower. 
Cyg X-2 has a larger disk (outer disk radius = 
3 $\times 10^{11}$ cm [Paper I])
than either Her X-1 (outer disk radius = $\sim 10^{11}$ cm [Cheng, Vrtilek, 
\& Raymond 1995]) or Sco X-1 (outer disk radius = 3$\times 10^{10}$ cm
[Vrtilek \etal~1991a]). 
Of the 10 models listed by Raymond (1993) none show O~V that do not 
{\it also} show N~IV] at a comparable or greater strength.
Raymond extrapolates from his models that the density will be about
$\sim 3\times 10^{13}$~cm$^{-3}$ at a radius of 3 $\times 10^{11}$ cm;
the lower density estimates of Paper I probably result
from the assumption that the N~V emission line is effectively optically
thin.
Kallman, Boroson, \& Vrtilek~(1998) point out that the O~IV~$\lambda$1340 
and O~V~$\lambda$1370 lines are subordinate lines, and
so require either that the emitting ions be in an excited state, or that
excitation occur from the ground state through a dipole-forbidden
transition.  The former case requires a combination of high gas density
and optical depth in the resonance line leading to the level from which
excitation can occur.  This can be expressed as $n\tau\ge10^{16}$
cm$^{-3}$.  Such densities and optical depths are predicted to occur in
X-ray heated disk atmospheres (Ko and Kallman 1994).

\section{Discussion and Conclusions}

As first demonstrated in Paper I,   
X-ray heating of a disk and companion star
provide good fits to the continuum ultraviolet
emission from Cyg~X-2/V1341~Cyg.  
We have refined the model described in Paper I by
using a distorted Roche-lobe filling surface for the optical
companion.
Owing to the
low temperature of
the companion star the contribution from the unheated star to 
the ultraviolet
flux is minor:
the primary contribution to the UV continuum is
X-ray heating effects on the disk.
We further changed the model in Paper I by distorting the
inner edge of the accretion into a saddle-shape
that reproduces the average X-ray lightcurve.
The reason for changing
only the inner disk are twofold:  the distortion is due to X-ray
heating effects and these are maximum in the inner disk; the disk
in Cyg X-2 is rather large (3 $\times 10^{11}$cm) and even small 
changes in the outer
disk would completely obscure the central X-ray emission. 
van Kerkwijk \etal~(1998) define a parameter F* that describes
where in a disk warping occurs.
When 0.1$<$F*$<$0.15, the outer disk
is warped, when 0.15$<$F*$<$0.2, the inner disk can be warped also,
and when F*$>$0.2, the inner disk can be tilted by $>$90 degrees
and may behave chaotically.  F* depends on the albedo of the
disk and the albedo can change as a function of angle of
incidence:  if we could prove that only warping of the inner
disk provides a solution than we have a limit on F* and from
that a limit on albedo.  We have not yet found a way to make
the model work by warping only the outer disk

The $\mdot$ values necessary to fit the simultaneous
GHRS/ASCA data we analyze here are
well within the range obtained using simultaneous IUE/EXOSAT data
of the system (Paper I).
We have increased these limits here in order to enclose ALL IUE
observations of the system.  
While the highest flux predicted by the model in the V band is 
greater than the highest
V magnitudes reported, the range in magnitude has the
same excursion as that observed.  The UV flux and U magnitude
measurements lie
well within the model predictions.

The line profiles of He II and N~V obtained with the G160M grating do
not show the double-peaked structure that is predicted by simple models
of line emission from disks.  This suggests that most of the line 
emission observed is from the surface of the companion; the measured
line velocities are consistent with this picture; however, the FWHMs
are larger than expected.  
It is also possible that the line emission is from an accretion heated
corona above the disk:  the relative strengths of the features observed
for Cygnus X-2 are consistent with the predictions of Raymond (1993);
however, the FWHMs are then smaller than expected.

The relatively
low-resolution GHRS observations are not able to
separate many line components that STIS will be able to resolve.
For example, if N~IV] is preferentially produced at the outer region 
of the disk 
 and
the line widths are due primarily to Keplerian velocities, 
we would  
expect it to be narrower than the other lines (the outer disk rotates
more slowly), that can also be formed in the inner disk. 
While there is some variation over the 
four G140L observations, the average values support this conclusion, with 
N~IV] having an average FWHM of 398 km~s$^{-1}$ whereas 
Si~IV~$\lambda$1393 averages
to 600~km~s$^{-1}$ and Si~IV~$\lambda$1402 averages to 459~km~s$^{-1}$ (this is
a $\sim$1$\sigma$ determination given our velocity resolution of 140km~s$^{-1}$).

The HST/STIS HST, extends in several significant
ways the
ultraviolet capabilities that were available with the GHRS:
with the echelle grating it is possible to sample continuously a broad
region (600$\ang$) of the spectrum at greater spectral
resolution than with the GHRS. 
STIS offers another improvement over the GHRS: a lower background
(the STIS dark count rate is
7.0~$\times 10^{-6}$~counts~s$^{-1}$~pixel$^{-1}$  
which is 50-100 times lower than that of the GHRS).  
For the GHRS, the background was not 
negligable compared with the source flux; in order of increasing
severity, this hampered the line, continuum, and variability
studies.
STIS observations will allow 
us to refine these as well as to 
investigate the physical condition of the emitting gas
through line and doublet ratios.

The X-ray spectra are well fit with the sum of a Comptonized bremmsstrahlung,
a blackbody, and an emission complex around 1 keV.
We note that if we associate the Comptonized component with the
neutron star and the blackbody+lines with emission from the
disk then the motion of the Z shape in the color-color diagram
could be attributed to the varying amounts of each component that
are covered by the disk.  This can be tested by looking at the
source during the high and the low states of the cycle:  the
motion of the Z should be correlated with the long-term period.
The correlation will not be absolute since the X-ray components also 
vary in strength depending
on mass accretion rate.  

SDV was supported by NASA through Grants NAG5-6711, and 
GO-07288.01-96A from the Space
Telescope Science Institute, which is operated by AURA, Inc., under NASA
contract NAS5-26555.

\newpage
\centerline{References}

Boroson, B., Vrtilek, S.D., McCray, R., Kallman, T.R., \& Nagase, F.
1996, ApJ, 473, 1079.

Boroson, B., Blair, W. P., Davidsen, A. F., Vrtilek, S.D.,
Raymond, J.C., Long, K. S., \& McCray, R. 1997, ApJ, 491, 903.

Branduardi-Raymont, G., Chiapetti, L., Ercan, E.N., 1984, A\&A,
130, 175.

Canizzo, J.K., \& Kenyon, S.J. 1987, ApJ, 320, 319

Cardelli, J. A., Savage, B. D., \& Ebbets, D. G. 1990, ApJ, 365, 789.

Cheng, F.H., Vrtilek, S.D., \& Raymond, J.C. 1995, ApJ, 452, 825

Chevalier, C., Bonazzola, S., \& Ilovaisky, S.A. 1976, A\&A, 53, 313. 

Chiapetti, L., Treves, A., Branduardi-Raymont, G.~\etal~1990, ApJ,
361, 596.

Clarkson, W.I., Charles, P.A., Coe, M.J., Laycock, S., Tout, M.D.,
\& Wilson, C.A. 2003, MNRAS, 339, 447 

Clarkson, W.I., Charles, P.A., Coe, M.J., \& Laycock, S. 2003,
MNRAS, in press, astro-ph 0304073.

Cowley, A.P., Crampton, D., \& Hutchings, J.B. 1979, ApJ, 231, 539.

Davies, S.R.  1990, MNRAS, 244, 93

Davies, S.R.  1991, MNRAS, 251, 64

Ebisawa, K., Ueda, Y., Inoue, H., Tanaka, Y., \& White, N. E.  1996,
ApJ, 467, 419. 

Hasinger, G., van der Klis, M., Ebisawa, K. Dotani, T., Mitsuda, K. 
1990, A\&A, 235, 131.

Howarth, I.D., \& Wilson, B. 1983a,  MNRAS, 204, 347.

Howarth, I.D., \& Wilson, B. 1983b,  MNRAS, 204, 1091.

Kallman, T.R., Boroson, B., \& Vrtilek, S.D. 1998, ApJ, 502, 441.

Kallman, T.R., \& Krolik, J. 1993, NASA Internal Report.

Kallman, T.R., \& McCray, R. 1982, ApJS, 50, 263.  

Ko, Y.K., \& Kallman, T.R. 1994, ApJ, 431, 273.
  
Kuulkers, E., Parmar, A.N., Owens, A., Oosterbroek, T., \& Lammers,
U. 1997, A\&A, 323, L29.

Kuulkers, E., van der Klis, M., Van Paradijs, J., 1995, ApJ,
450, 748.

Lang, K.R. 1980, Astrophysical Formulae, 
Springer-Verlag: New York.

Levine, A.M., \etal~1996, ApJ, 469, L33.

Lum, K.L., Canizares, C.R., Clark, G.W.,~\etal~1992, ApJ, 78, 423.

McClintock, J.E., Petro, L.D., Hammerschlag-Hensberge, G., Proffitt, C.R.,
\& Remillard, R.A. 1984, ApJ, 283, 794

Morrison, R., \& McCammon, D. 1983, ApJ, 270, 119.

Paul, B., Kitamoto, S., \& Makino, F. 2000, ApJ, 528, 410

Psaltis, D., Lamb, F.K., and Miller, G.  1995, ApJ, 454, L137

Raymond, J.C. 1993, ApJ, 412, 267.

Smale, A. P., Angelini, L., White, N. E., Mitsuda, K., \& Dotani, T.
1994, BAAS, 185, 102.
 
Smale, A.P., Done, C., Mushotzky, R.F.,~\etal~1993, ApJ, 410, 796.

Sunyaev, R.A. \& Titarchuk, L.G. 1980, A\&A, 86, 121.

Tanaka, Y., Inoue, H., \& Holt, S.S. 1994, PASJ, 46, L37.

van Kerkwijk, M.H., Chakrabarty, D.,  Pringle, J.E., \&
Wijers, R.A.M.J. 1998, ApJ, 499, L27. 

Vrtilek, S.D., Boroson, B., Cheng, F.H., McCray, R., \& Nagase, F.
1997, ApJ, 490, 377. 

Vrtilek, S.D., Kahn, S.M., Grindlay, J.E., Helfand, D.J., \&
Seward, F.D. 1986, ApJ, 307, 698.

Vrtilek, S.D., McClintock, J.E., Seward, F.D., Kahn, S.M.,
\& Wargelin, B.J. 1991a, ApJS, 76, 1127. 

Vrtilek, S.D., Penninx, W., Raymond, J.C., Verbunt, F.,
Hertz, P., Wood, K., Lewin, W.H.G., \& Mitsuda, K. 1991b, ApJ, 376, 278.

Vrtilek, S.D., Raymond, J.C., Garcia, M.R.,
Verbunt, F., Hasinger, G., \& Kurster, M. 1990, A\&A
235, 162 (Paper I). 

Vrtilek, S.D., Swank, J.H., Kallman, T.R. 1988, ApJ, 326, 186.

Wijnands, R.A.D., Kuulkers, E., \& Smale, A.P. 1996, ApJ, 473, L45.

Wijnands, R.A.D., van der Klis, M., Kuulkers, E., \& Hasinger, G.
1997, A\&A, 323, 399. 

\newpage
\centerline{Figure Captions}

\noindent
{\bf Figure 1.}
(a) One day averages of the flux observed from Cyg X-2 with the All Sky
Monitor on board the ROSSI X-ray Timing Explorer.  (Quick-look results
were provided by the ASM/RXTE team.) (b) The data from (a) (histograms)
with the lightcurve from (c) (smooth curve) superposed.
(c) The lightcurve obtained by folding all RXTE/ASM  
data with a period of  
81.7 days).  In (c) the arrow indicates the time of
the simultaneous HST/ASCA observations.
\vskip 0.1in
\noindent
{\bf Figure 2.}
X-ray lightcurves (0.6-10 keV) from 1995 Dec. 6 22:15 UT to 1995 
Dec. 7 22:01 UT
(from SIS1 on ASCA) and ultraviolet coverage
(with GHRS on HST).  The numbers A1-A11 refer to ASCA orbit.  The numbers
H1-H8 refer to HST observation.  The GHRS filters and wavelength ranges
are indicated.
\vskip 0.1in
\noindent
{\bf Figure 3.}
HST/GHRS G160M observations.
H1 and H2 are centered on N~V.  H3 and H4 are centered on HeII.
The thin solid lines show the best fit to the continuum using the
model described in Section 3. 
\vskip 0.1in
\noindent
{\bf Figure 4.}
HST/GHRS G140L observations.  The thin solid lines show the best fit
to the continuum using the
model described in Section 3. 
\vskip 0.1in
\noindent
{\bf Figure 5.}
An X-ray color-color diagram for the ASCA data.
The ratio taken is that of photon counts uncorrected for effective area.
A schematic ``Z-curve'' is drawn on for comparison.
\vskip 0.1in
\noindent
{\bf Figure 6.}
The saddle-shape used for the 
the inner edge of the disk.
Lines-of-sight to the observer at long-term phases of X-ray maxima
(0.25 and 0.75) and X-ray minima (0.5 and 0.0) are indicated. 
\vskip 0.1in
\noindent
{\bf Figure 7.}
Histograms represent the RXTE ASM data. The X-ray flux 
predicted from the
saddle-shaped disk is depicted as a dotted line.
The dashed line is a sinusoidal fit (fundamental plus
1st harmonic): 
a sin (2$\pi$(t-t$_o$)/82days) + b(2$\pi$(t-t$_o$)/41 days).
Fits of the saddle model to the data give a $\chi_{\nu}^2$ of 1.8 
whereas the fit using 
2 sinusoids gives a $\chi_{\nu}^2$ of 3.2. per degree of freedom.
\vskip 0.1in
\noindent
{\bf Figure 8.}
In all three panels the dotted lines represent X-ray low-states  
and dot-dashed lines represent X-ray high states of the 82-day period.
The upper curves are computed for the highest $\mdot$ and the lower
curves the lowest $\mdot$ required to include the observed 
UV fluxes. 
(a) Shows average UV continuum flux prediced by the model near N~V vs. 
binary phase.  
The solid
triangle represents the GHRS observations and the open triangles are 
archival
IUE data.  
(b) The V
magnitude predictions of the model (average over 5000-6000$\AA$).
The dashed lines denote the extremes of V observed for the system.
(c) The U magnitude predictions of the model
(average over 3510-3520$\AA$; The IUE flux library used for our
model have no values between 3200-3500$\AA$).
The dotted lines represent extremes of U magnitudes reported for Cyg X-2.
\vskip 0.1in
\noindent
{\bf Figure 9.}
Histograms represent data shown as unfolded photon spectra.
Typical error bars are depicted in each panel. 
(a) ASCA GIS 3 spectrum for orbits A1-A4.
(b) ASCA GIS 3 spectrum for orbits A6-A8.
(c) ASCA GIS 3 spectrum for orbits A9-A11.
(d) ASCA GIS 3 spectrum for the PV phase observation.
The solid lines represent the best-fit models as
listed in Table 2.
The dotted lines show the Comptonized Bremmsstrahlung component; the long dashes 
the blackbody
component; and the short dashes the 
line complex near 1 keV.
\vskip 0.1in
\noindent
{\bf Figure 10.}
An HST G160M observation of the N~V doublet.  The narrow lines have
velocities of -93 and -94 km~s$^{-1}$ with FWHMs of 316 and 
336 km~s$^{-1}$.  The
broad lines (not listed in Table 3) have
velocities of -210~km~s$^{-1}$ with FWHM of 1000~km~s$^{-1}$.
\vskip 0.1in
\noindent
{\bf Figure 11.}
A comparison of HST GHRS 140L observations of Cyg X-2, Sco X-1, and
Her X-1.  The solid
lines show the best-fit to the continuum using X-ray heated
disk and companion star. 
The corresponding mass-accretion rates are noted.
\small
\singlespace
\newpage
\topmargin -0.5in

\centerline{\sc Table 1. Parameters Used for Ultraviolet Continuum Fits$^1$}
\begin{center}
\vskip 0.1in
\begin{tabular}{lc} \hline \hline
&\\
Parameter & Value\\
&\\
\hline
&\\
Mass of neutron star M&1.4$\msun$\\ 
&\\
Mass ratio $q = (M_{NS}/M_{star})$&2.0\\
&\\
Temperature of normal star&7000K\\
&\\
E(B-V)&0.45\\
&\\
Distance to source&8.0 kpc\\
&\\
Separation {\it a}&1.8$\times 10^{12}$cm\\
&\\
Inner disk radius {\it r$_1$}& 1.0$\times 10^6$cm\\
&\\
Outer disk radius {\it r$_2$}&3.0$\times 10^{11}$cm\\
&\\
Gravity darkening exponent $\beta$ &0.08\\
&\\
Limb darkening coefficient&0.8\\
&\\
Inclination angle of orbital plane {\it i}&73.0 deg\\
&\\
Radius of maximum warp R$_{mw}$&1.7 $\times$ r$_{NS}$\\
&\\
Warp opening angle $\theta_w$&40.8$\pm$1.5 deg\\
&\\
Disk angular semi-thickness $\theta_d$&8.0 deg\\
&\\
Tilt of inner disk to orbital plane $\alpha_{d}$&4.1$\pm$0.4 deg\\ 
&\\
Line-of-nodes parameter $\Delta \psi$ &2.95$\pm$0.04\\
&\\
&\\ \hline
\end{tabular}
\end{center}
$^1$1$\sigma$ errors are given for fitted parameters.
Parameters without errors are as taken from Cowley, Cramption,
\& Hutchings 1979, McClintock~\etal~1984, and Paper I.

\newpage
\topmargin -0.5in

\centerline{\sc Table 2. ASCA (GIS3) Spectral Fits (0.6 - 10 keV)$^1$}
\begin{center}
\vskip 0.1in
\begin{tabular}{lcccc} \hline \hline
&&&&\\
Parameter & A1-4$^2$ & A6-8$^2$ & A9-11$^2$ & PV Phase$^3$\\
&&&&\\
\hline
Exposure time (s)&6,592&6,010&6,272&22,703\\
&&&&\\
Count rate (cts s$^{-1}$)&84.6$\pm$0.1&81.4$\pm$0.1&85.8$\pm$0.1&84.9$\pm$0.1\\
&&&&\\
N$_{H}$$^4$ (10$^{21}$cm$^2$) & 0.7$\pm$0.5 & 0.1$\pm0.4$ & 0.5$\pm0.1$ & 1.8$\pm0.2$\\
&&&&\\
CompST$^5$ kT (keV) & 2.1$\pm0.1$ & 2.1$\pm0.1$ & 2.0$\pm0.1$ & 1.6$\pm0.02$\\
&&&&\\
CompST $\tau$ & 25$\pm4$ & 29$\pm5$ & 27$\pm6$ & 23$\pm1$\\ 
&&&&\\
Blackbody$^6$ kT (keV) & 0.54$\pm0.01$ & 0.55$\pm0.02$ & 0.60$\pm0.01$ & 
0.62$\pm0.02$\\ 
&&&&\\
Fraction of energy flux from blackbody&0.16&0.20&0.10&0.10\\
&&&\\
Line$^7$ energy (keV) & 1.02$\pm0.2$ & 1.09$\pm0.4$ & 0.95$\pm0.3$ 
& 1.02$\pm0.3$\\
&&&&\\
Line width (keV) & 0.2$\pm0.2$ & 0.3$\pm0.2$ & 
0.2$\pm0.2$ & 0.11$\pm0.02$\\ 
&&&&\\
Eq. width of line (eV) & 66  & 75  & 94 & 38\\
&&&&\\
Reduced $\chi^2$ & 1.4 & 1.2 & 1.4 & 1.8\\
&&&\\ \hline
\end{tabular}
\end{center}
$^1$All errors given are 1 $\sigma$.\\
$^2$ASCA observation identifier: 43029000.\\ 
$^3$ASCA observation identifier: 40017000.\\
$^4$N$_H$ is the neutral hydrogen column density.
The effective absorption cross section $\sigma$(E) was calculated
on the assumption of neutral cosmic-abundance material (Morrison and
McCammon 1983).\\ 
$^5$Comptonization model of Sunyaev \& Titarchuk (1980; Comptonization
of cool photons on hot electrons).\\ 
$^6$N(E)(cm$^{-2}$~s$^{-1}$~keV$^{-1}$) = A$_{BB}$(e$^{E/kT}$-1)E$^2$\\ 
$^7$A simple Gaussian.  \\
N(E)(cm$^{-2}$~s$^{-1}$~keV$^{-1}$) = A$_{Line}(1/(\sqrt{2\pi\sigma^2})e^{(-0.5(E-E_{Line})^2/\sigma^2)}$\\ 
E is energy in keV; T is temperature in Kelvin, $k$ is the Boltzmann 
constant, A$_{BB}$ and A$_{Line}$ are normalization constants. 
\newpage
\begin{center}
{\bf 
{\sc Table 3. Ultraviolet Line Fluxes and Velocities }
}
\vskip 0.05in
\begin{tabular}{lcccc} \hline \hline
HST&&Flux&FWHM&Velocity$^1$\\ 
OBS No.&Feature&(erg~s$^{-1}$~cm$^{-2})$&(km~s$^{-1}$)&(km s$^{-1}$)\\
\hline \\
&&&& \\
H1$^2$ & N V 1238& 5.9$\pm$0.8(-14) & 272$\pm$25&-99$\pm$10 \\
G160M&        N V 1242& 3.8$\pm$0.7(-14)&239$\pm$31& -92$\pm$13\\
&&&&\\
H2 &	N V 1238& 4.8$\pm$0.4(-14) &336$\pm$23& -94$\pm$9 \\
G160M&        N V 1242& 3.9$\pm$0.4(-14)& 316$\pm$24&-93$\pm$9\\ 
&&&& \\
H3  &     He II 1640& 3.1$\pm$0.6(-14)&343$\pm$46&-145$\pm$20\\ 
G160M&&&&\\
&&&&\\
H4 &     He II 1640& 2.8$\pm$0.6(-14)&327$\pm$46&-140$\pm$29\\ 
G160M&&&&\\
&&&& \\
H5 &      1290-1550$\ang$&8.87(-13)&...&...\\ 
G140L&Si IV 1394& 1.3$\pm$0.2(-14)&580$\pm$63&-162$\pm$26\\ 
&Si IV 1403& 0.9$\pm$0.2(-14)&366$\pm$49&-70$\pm$22\\  
  &     N IV] 1486& 1.1$\pm$0.2(-14)&412$\pm$55&-179$\pm$25\\ 
  &     C IV 1548& $>$7.8(-15)&...&-80 (at peak)\\ 
&&&& \\
H6 &     1290-1550$\ang$&8.89(-13)&...&...\\ 
G140L&Si IV 1394& 1.2$\pm$0.2(-14)&591$\pm$69&-178$\pm$29\\ 
  &     Si IV 1403& 0.9$\pm$0.2(-14)&426$\pm$57&-150$\pm$25 \\
  &     N IV] 1486& 1.3$\pm$0.2(-14)&492$\pm$66&-128$\pm$28\\ 
  &     C IV 1548&$>$5.5(-15)&...&-140 (at peak)\\ 
&&&& \\
H7 &     1290-1550$\ang$&9.36(-13)&...&...\\ 
G140L  &     Si IV 1394& 1.4$\pm$0.2(-14)&586$\pm$61&-175$\pm$26\\ 
  &     Si IV 1403& 1.2$\pm$0.2(-14)&567$\pm$67&-150$\pm$29\\ 
  &     N IV] 1486& 1.2$\pm$0.2(-14)&392$\pm$53&-162$\pm$23\\ 
&C IV 1548&  $>$7.8(-15)&...&-140 (at peak)\\ 
&&&& \\
H8 &     1290-1550$\ang$&8.66(-13)&...&...\\ 
G140L&Si IV 1394& 1.4$\pm$0.2(-14)&642$\pm$63&-182$\pm$29\\ 
&Si IV 1403& 1.3$\pm$0.2(-14)&476$\pm$47&-148$\pm$21\\ 
  &     N IV] 1486& 1.0$\pm$0.2(-14)&296$\pm$34&-148$\pm$15\\ 
  &     C IV 1548& $>$1.0(-14)&... &-100 (at peak)\\       
&&&& \\\hline \hline \\
\end{tabular}
\end{center}
$^1$Heliocentric.  Heliocentric velocity of center-of-mass of
the Cyg X-2 system is -220 km~s$^{-1}$ as given by
Crampton, Cowley, \& Hutchings (1979).\\ 
$^2$The ``H'' numbers are the same as in Figure 2.  The HST 
observation identifications are:  Z31U01\#T where \# is 4 for
H1; 6 for H2; 8 for H3 and Z3U010\#T where \# is A for H4; C for H5; 
E for H6; G for H7; and I for H8.  Alternating number observations
are SPYBAL exposures used for calibration.
The line velocities have been corrected for the SPYBAL exposures.

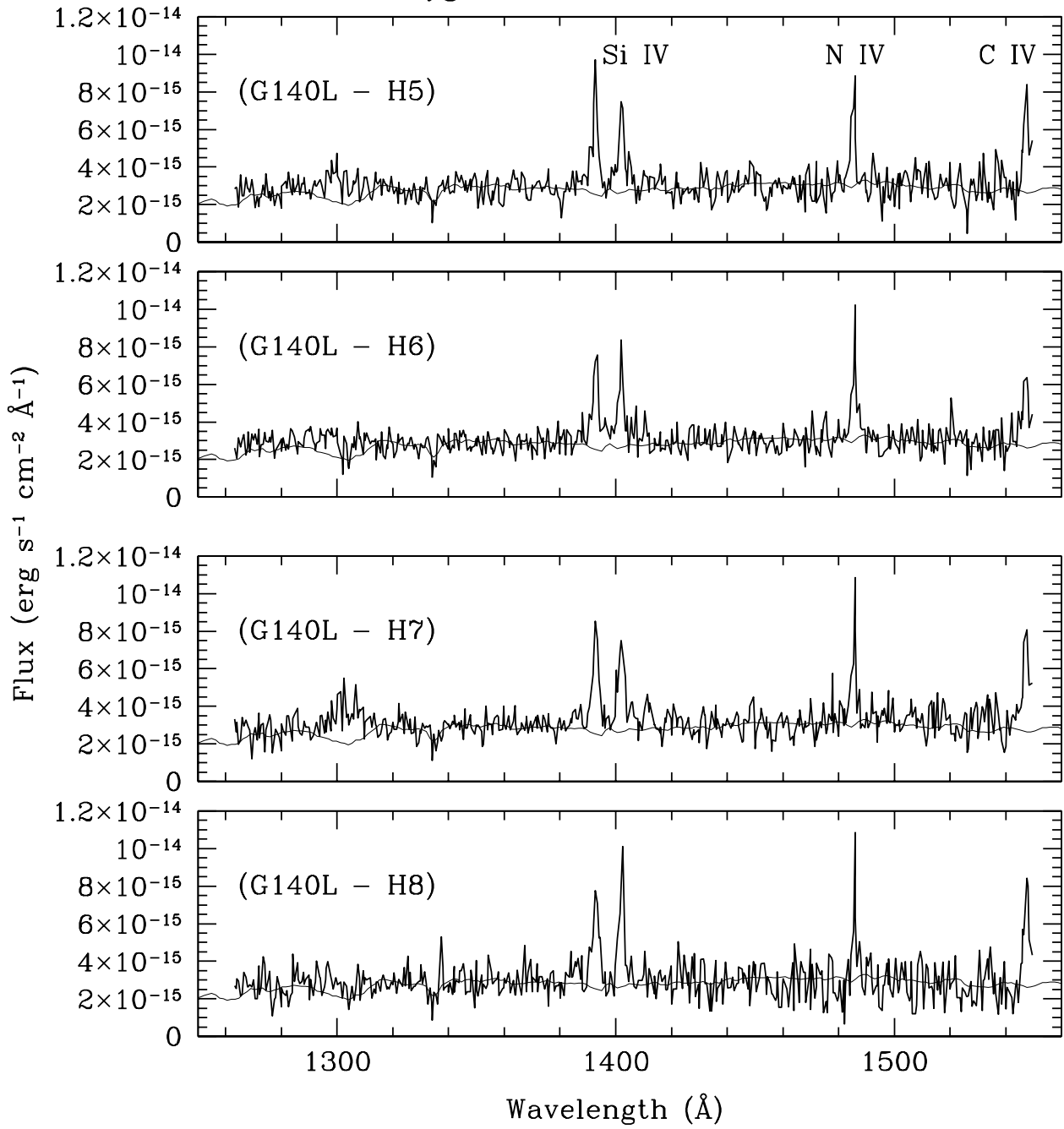
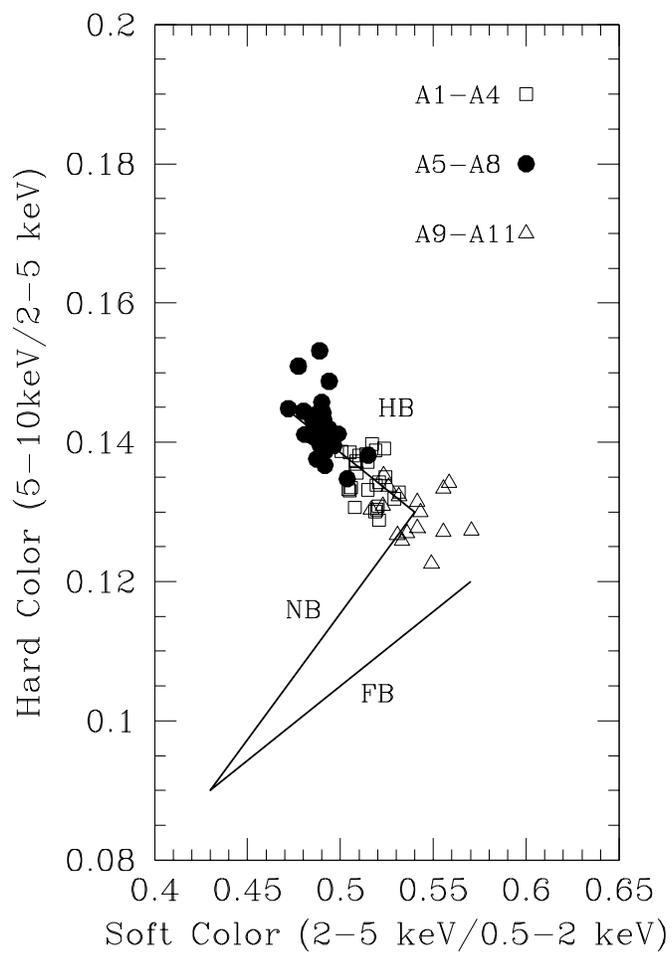
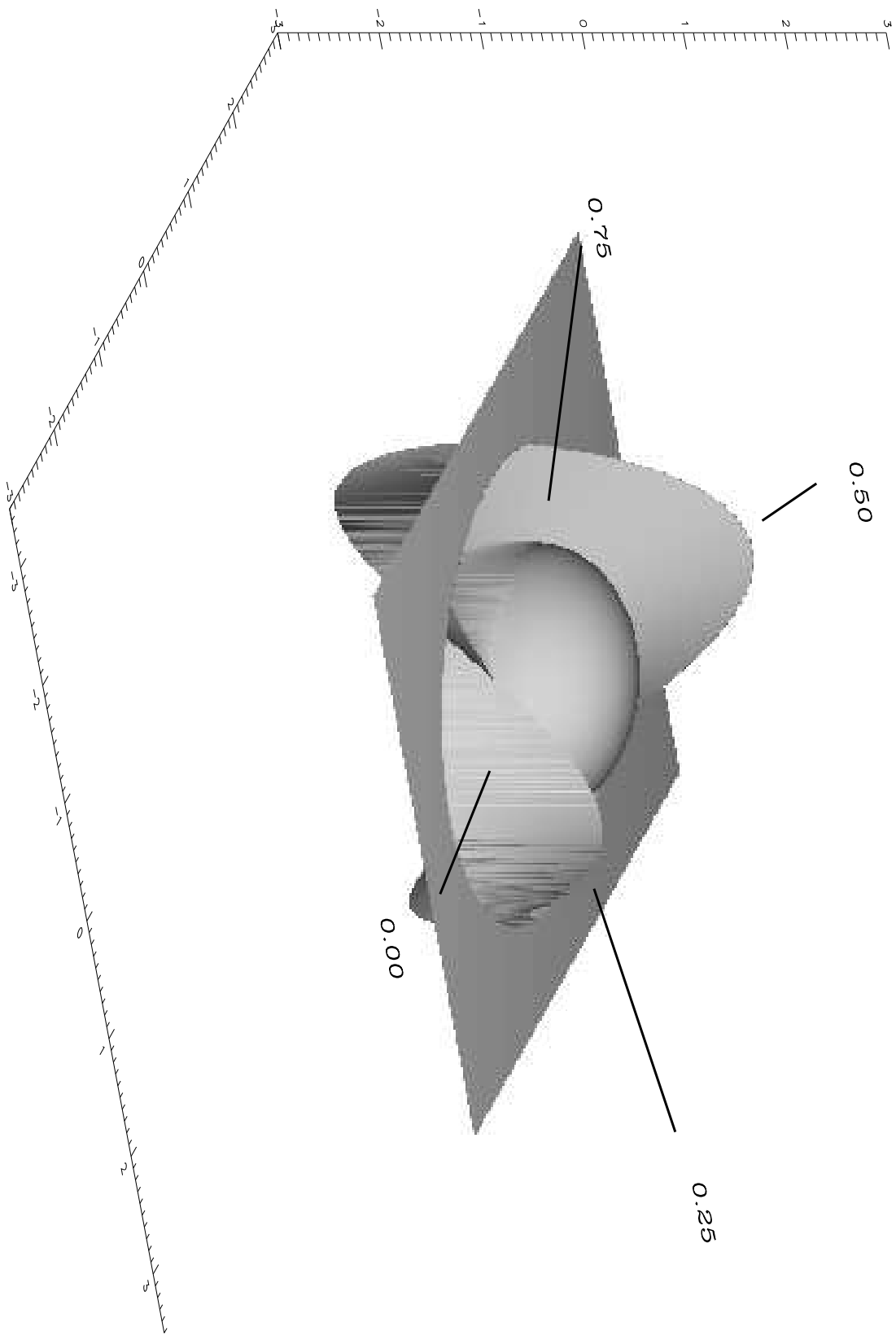
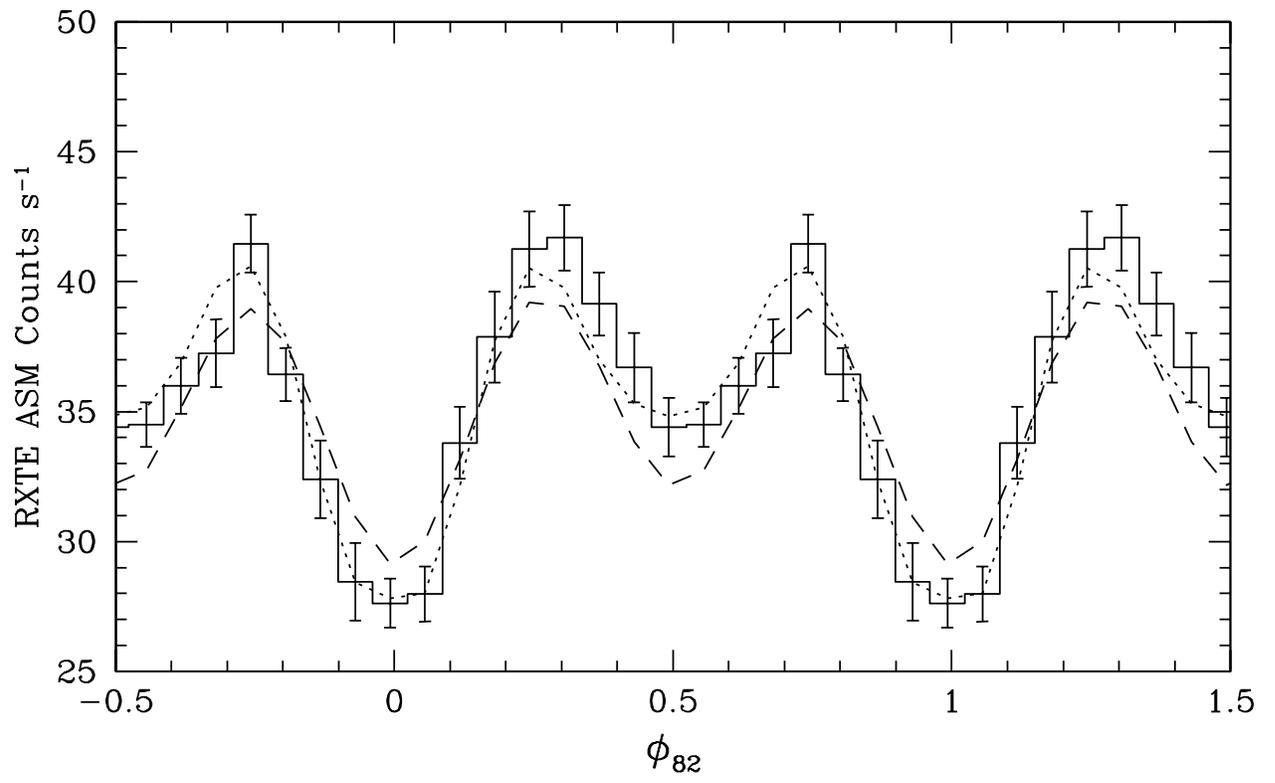
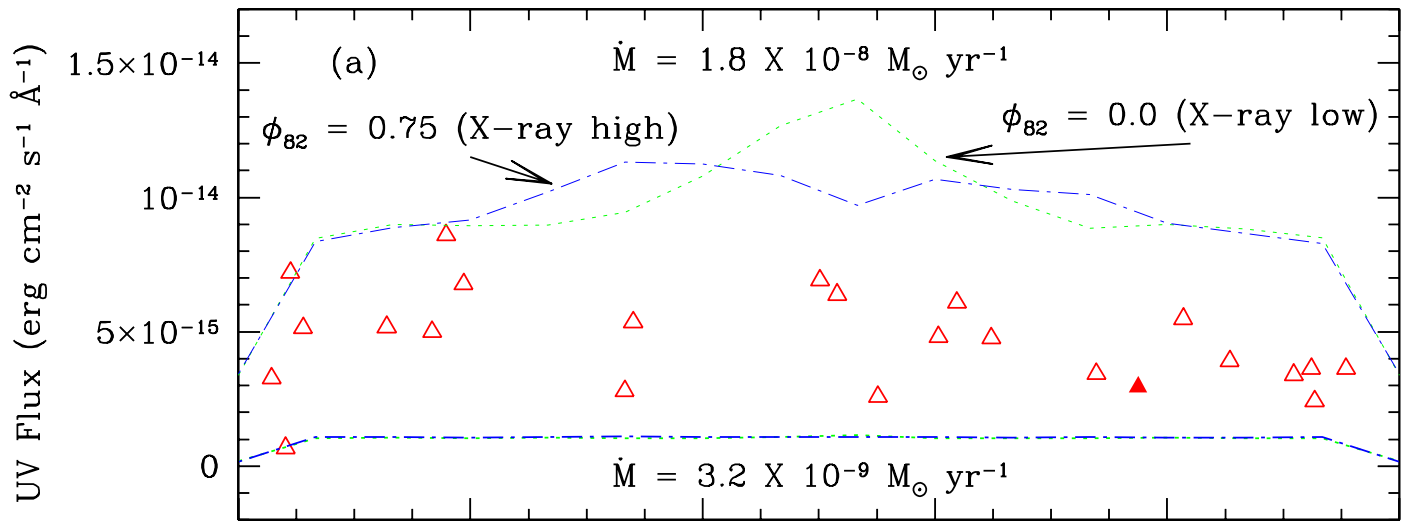
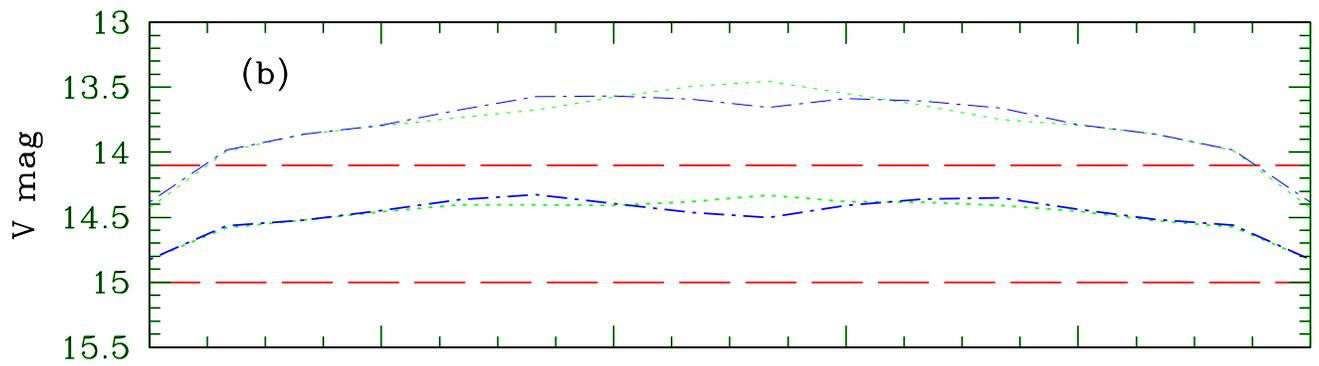
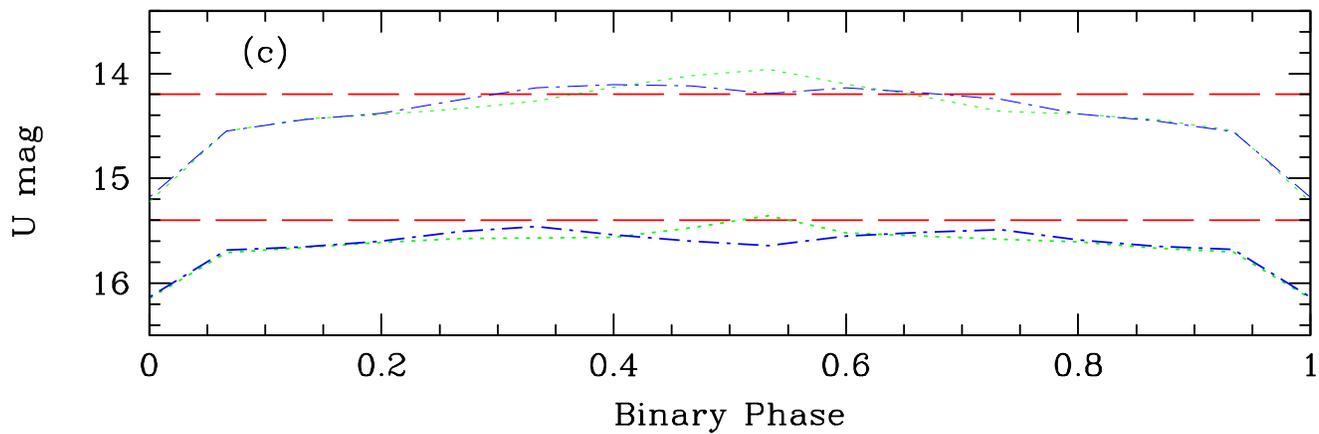
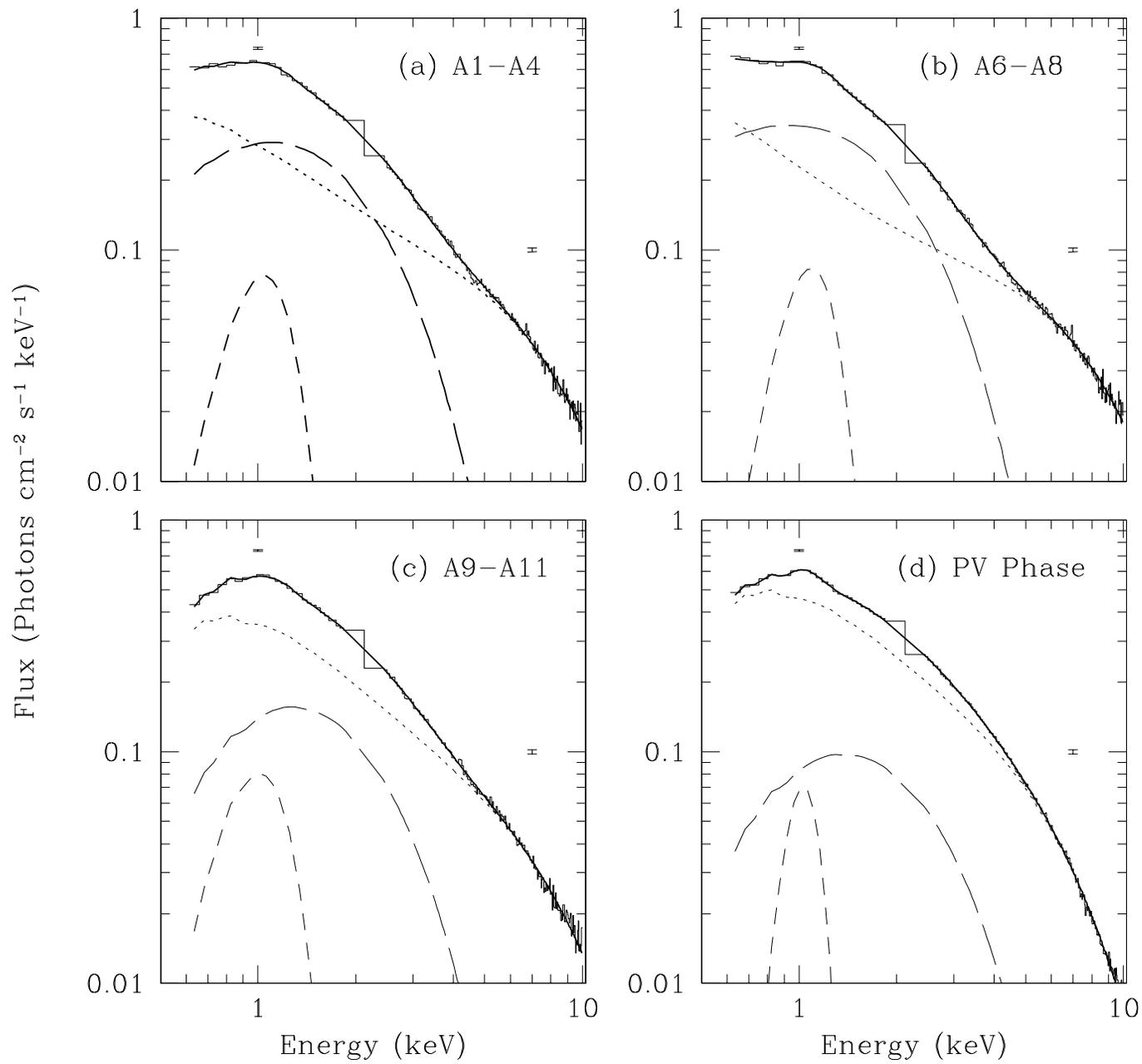
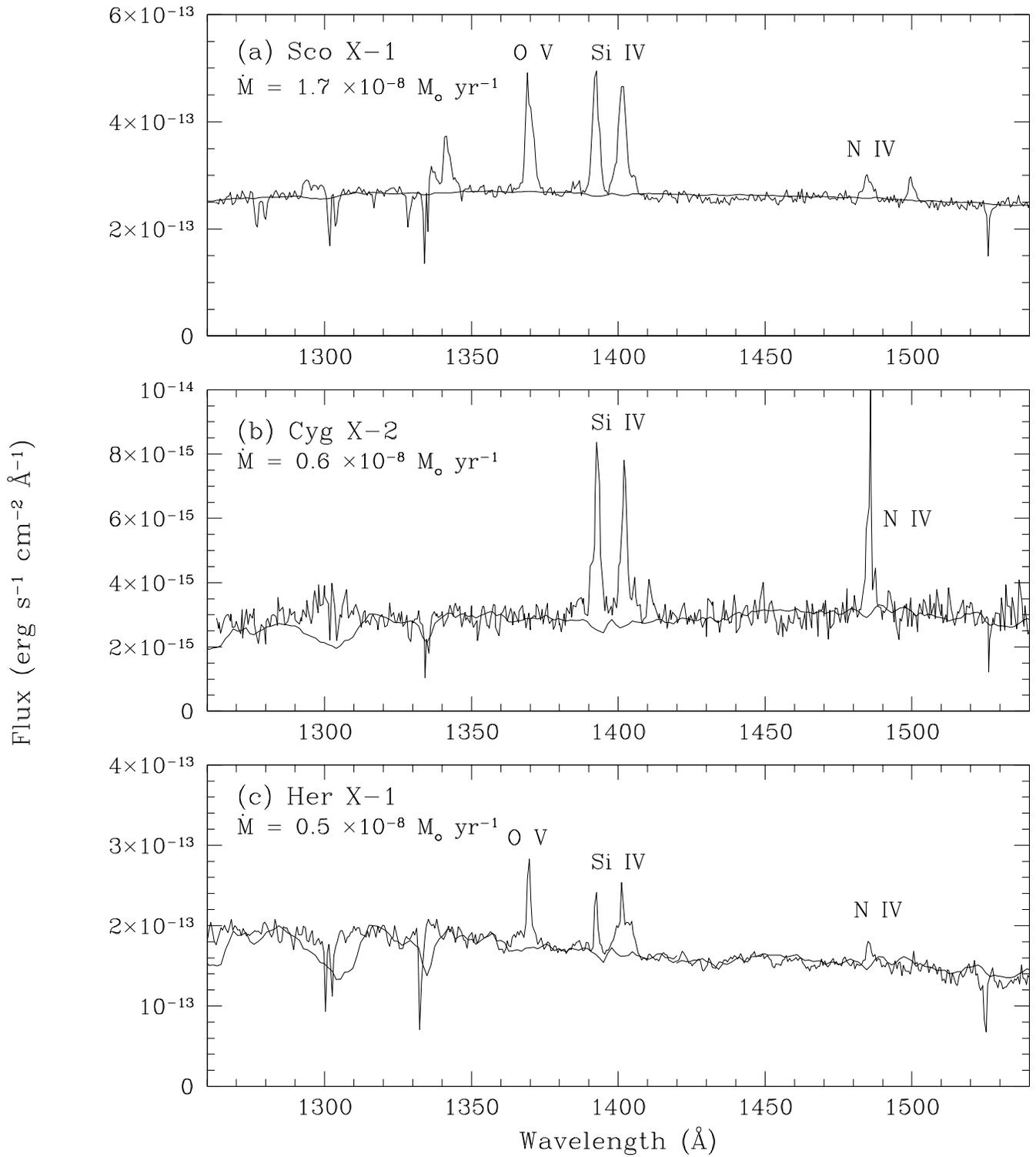
\end{document}